\documentstyle[prl,twocolumn,aps]{revtex} \input psfig
\begin{document}
\draft
\title{
Persistent current in a one-dimensional ring 
of fractionally charged ``exclusons'' }
\author{I. V. Krive$^{(1,2)}$, P. Sandstr{\"o}m$^{(1)}$, R. I. Shekhter$^{(1)}$, and M. Jonson$^{(1)}$ } 
\address{$^{(1)}$Department of Applied Physics, 
Chalmers University of Technology and G{\"o}teborg University, 
S-412 96 G{\"o}teborg, Sweden\\
and $^{(2)}$B. Verkin Institute for Low Temperature Physics and Engineering, 
Academy of Sciences of Ukraine, 310164 Kharkov, Ukraine}
\maketitle 
\begin{abstract}
The Aharonov-Bohm effect in a one-dimensional (1D) ring containing a gas of
fractionally charged excitations is considered. It is shown that the low
temperature behavior of the system is identical to that of free electrons with
(integer) charge $e$. This is  a direct consequence of the fact that the total
charge in the ring is  quantized in units of the electron charge. Anomalous 
oscillations of the persistent current amplitude with temperature are predicted
to occur as  a direct manifistation of the fractional nature of the quasiparticle
charge. A 1D conducting ring with gate induced periodical potential is discussed
as a possible set-up for an experimental observation of the predicted
phenomenon.
\end{abstract}

One of the most intriguing predictions of low dimensional models
in quantum field theory and in condensed matter physics is the existence
of fractional charge.
(See, e.g., the reviews \onlinecite{Niemi} and \onlinecite{Krive.SovPhys}).
The necessary condition for fractional charge quantization is that the charge
carrying quasiparticles are topologically stable. In 1D models the stable
quasiparticles as a rule are kinks of the order parameter, while in 2D
models they are vortices. Nontrivial topological properties of fractionally
charged quasiparticles are responsible for the unusual statistics of these
objects. Even in the absence of interactions (ideal gas) the quantum ensemble
obeys statistics which is intermediate between traditional Bose-Einstein and
Fermi-Dirac statistics.

Some years ago, Haldane \onlinecite{Haldane} suggested a generalization
of Pauli's exclusion principle to consider anomalous statistics by
standard methods of quantum statistical mechanics. Such an approach was
realized in Ref.~\onlinecite{Wu}, where the thermodynamic properties 
of a generalized ideal gas were calculated. 
In this paper we will study
coherent properties of mesoscopic rings characterized by quasiparticles 
carrying fractional charge and obeying fractional exclusion statistics. 
These stable quasiparticles will be referred to as exclusons.
We calculate the persistent current of an ideal one-dimensional ring of 
non-interacting exclusons at finite temperatures by using two approaches, viz.
%
(i) by
   a direct method involving the distribution function of exclusons
   \onlinecite{Wu} (see also Ref.~\onlinecite{Dasnieres})
and (ii) by
   mapping the excluson gas model to the Luttinger liquid model
   \onlinecite{WuYu}.
We will show that at low temperatures $T \ll \Delta$
(where $\Delta$ is the energy level spacing of a ballistic ring), the persistent
current of fractionally charged exclusons in an isolated ring or in a ring
weakly connected to an electron reservoir is identical to the persistent
current of non-interacting electrons carrying the same total electric charge.
It follows that anomalous statistics does not show up in 
Aharonov-Bohm oscillations of a perfect (impurity free) integrable
charged system.
This result (for $T=0$) also follows from a general theorem \onlinecite{Muller}.

The high temperature properties of the persistent current of exclusons differ,
however, qualitatively from those of non-interacting particles with normal
statistics;  we find an anomalous temperature dependence of the amplitude of the
persistent current (the amplitude oscillates as a function of temperature). We
therefore propose that the effects predicted here should be observable in
experiments using loop-shaped quantum wires  with a gate induced periodic
potential. They should be revealed through the equilibrium properties close to
the  commensurate-incommensurate phase transition.  


Let us consider a noninteracting gas of (spinless) exclusons with spectrum
$\epsilon(p)$ and fractional charge $q_{s}=e/q$ ($e$ is the electron
charge, $q$ is an integer). It is known that in non-simply connected regions,
the wavefunction of fractionally charged quasiparticles is
multi-valued \onlinecite{Kivelson,Bogachek.JETP}. 
As an illustration of the above assertion, we consider kinks (solitons)
of the order parameter in a system undergoing a 
commensurate-incommensurate phase transition. In the commensurate
(dielectric) phase, the existence of kinks with fractional charge
$q_{s}=e/q$ is related to the existence of $q$ degenerate vacua, $|0_{k}\rangle$
($k = 1,.\ldots,q$). The transport of a given soliton along the ring transforms
the vacuum $|0_{k}\rangle$ to the vacuum $|0_{k \pm 1}\rangle$ (the $\pm$-signs
correspond to the two directions of motion) and only a $q$-fold
encircling of the ring restores the initial vacuum state. Therefore the boundary
condition for the wave function of the topological defects (kinks) in a ring
takes the form \onlinecite{Kivelson} $\Psi (x+qL)=\Psi(x)$. In the interval
[$0,L$] the wave function is multi-valued and can be defined on $q$ sheets.
On the $k$-th sheet one has
\begin{equation}
\Psi_{k}(x)=\exp \left(i \frac{2 \pi k}{q} \frac{x}{L} \right)
\Psi_{per}(x), \quad k=1,2,..,q
\label{periodic1}
\end{equation}
In the presence of a magnetic flux, $\Phi$, the periodic boundary condition
for $\Psi_{per}(x)$ has to be changed to a twisted one
\begin{equation}
\Psi_{per}(x+L) = \exp \left( i 2 \pi \frac{\Phi}{\Phi_{s}} \right)  \Psi_{per}(x) ,
\label{periodic2}
\end{equation}
where $\Phi_{s} = hc/q_{s} = q\Phi_{0}$ ($\Phi_{0} = hc/e$ is the flux
quantum). Using Eqs.~(\ref{periodic1}) and (\ref{periodic2}) one gets the
quantized spectrum for fractionally charged particles in a ring (see also
\onlinecite{Bogachek.JETP}) 
\begin{equation}
\epsilon_{n,k} = \epsilon \left( n + \frac {k}{q} + \frac{1}{q} 
\frac{\Phi}{\Phi_{0}} \right), \quad k=1,..,q
\label{spectrum}
\end{equation}
where $n$ is an integer. The index $k$ reflects the fact
that the wavefunction is multi-valued. The spectrum (\ref{spectrum})
describes $q$ branches and for any equilibrium state of fractionally charged
quasiparticles one has to sum over $k$ from 1 to $q$.

As a consequence of gauge invariance fractionally charged quasiparticles 
in condensed matter systems obey fractional statistics 
(see, e.g., Ref.~\onlinecite{Hatsugai}). 
We will regard them as exclusons with statistical
parameter $\lambda = 1/q$ ($q=1$, $\lambda$=1 corresponds to fermions,
$\lambda$=0 --- to bosons). The thermodynamic potential, $\Omega$, for
an ideal gas of exclusons is of the form \onlinecite{Wu}
\begin{equation}
\Omega_{\lambda} = - \frac{1}{\beta} \sum_{\{ p \}} \ln \left[
\frac{1+(1-\lambda)f_{p}}{1-\lambda f_{p}} \right] ,
\label{potential}
\end{equation}
where $\beta$ = T$^{-1}$ is the inverse temperature. The one-particle 
distribution function, $f_{p}$, in Eq. (\ref{potential}) takes the form
\begin{equation}
f_{p}=\left[y(\xi) + \lambda \right]^{-1}, \quad \xi = \exp \left[
\beta(\epsilon_{p} - \mu_{s}) \right],
\label{f}
\end{equation}
where $\epsilon_{p}$ is the one-particle spectrum, $\mu_{s}$ is the chemical
potential and the dimensionless function $y(\xi)$ obeys the algebraic
equation
\begin{equation}
y^{\lambda}(1+y)^{1-\lambda} = \xi .
\label{y}
\end{equation}
With the help of Eqs. (\ref{potential})-(\ref{y}) the general expression
for the persistent current $I = -c \partial \Omega_{q} / \partial \Phi$
of exclusons can be written as
\begin{equation}
I_{q}(\Phi) = -c \sum_{n=-\infty}^{\infty} \sum_{k=1}^{q}
f_{n,k} \frac{\partial \epsilon_{n,k}}{\partial \Phi} .
\label{current}
\end{equation}
In what follows we will for simplicity consider a linearized spectrum
(around the Fermi momentum), of Eq. (\ref{spectrum})
\begin{equation}
\epsilon_{n,k} = \Delta |n + \frac{k}{q} + \frac{1}{q}
\frac{\Phi}{\Phi_{0}}| ,
\label{linspectrum}
\end{equation}
where $\Delta \equiv 2 \pi \hbar v_{F} / L$ is the energy level spacing.
Notice that the persistent current being a property of the Fermi surface
\onlinecite{Cheung} (see also the review \onlinecite{Zwyagin}),
does not depend on the energy dispersion far from the Fermi level. 
Next we will assume that the total charge of the ring is an integer
multiple of $e$ (this is valid for an isolated ring or for a ring weakly
connected to a reservoir of electrons).
In this case the number of exclusons in a ring is a multiple of $q$.
By taking into account the fact that at $T=0$ the distribution function, $f_{p}$,
is a step of height $q$ \onlinecite{Wu}, we can infer that the
Fermi velocity entering the definition of $\Delta$ does not depend on
the statistical factor $q$. 
When substituting Eqs.~(\ref{f}) and (\ref{linspectrum}) into the equation
for the persistent current $I_{q}(\Phi)$ and performing the summation over $n$
in  Eq.~(\ref{current}) using the Poisson summation formula,
we get the following expression
\begin{eqnarray}
I_{q}(\Phi) = 2 \frac{e T}{\hbar} \sum_{j=1}^{\infty} 
\sin \left( 2 \pi j \frac{\Phi}{\Phi_{0}} \right) &
\left\{
S_{j}^{(q)} \cos \left( 2 \pi j \frac{q \mu_{s}}{\Delta} \right) + 
\right. \nonumber \\ & \left.
C_{j}^{(q)} \sin \left( 2 \pi j \frac{q \mu_{s}}{\Delta} \right) \right\} ,
\label{current2}
\end{eqnarray}
where
\begin{equation}
\left\{ \begin{array}{ll}
S_{j}^{(q)}  \\ 
C_{j}^{(q)} \end{array} 
\right\} =
- \frac{1}{\pi} \int_{- \beta \mu_{s}}^{\infty} 
\frac{dx \left\{ \begin{array}{ll} \sin \\ \cos \end{array} \right\}
\left(2 \pi j q \frac{x}{\beta \Delta} \right) } {y(e^{x}) + 1/q}
\label{SC}
\end{equation}
and the function $y(\xi)$ satisfies Eq. (\ref{y}). 
Since we are interested in the temperature region 
$T \ll \mu_{s}$, the lower limit in the
integrals of equation (\ref{SC}) can be replaced by -$\infty$.
For the limit of Fermi statistics ($q=1$) it then follows that
\begin{equation}
S_{j}^{(1)} = 1/\sinh \left( j \frac{2 \pi^{2}}{\beta \Delta} \right) ,
\quad C_{j}^{(1)} = 0 .
\label{SC2}
\end{equation}
In fact the integrals $C_j^{(1)}$ are exponentially small when the parameter 
$\beta \mu \gg 1$ and one recovers the well-known expression for the persistent
current of non-interacting (spinless) fermions
\onlinecite{Kulik.BIL,Cheung}.

From Eq. (\ref{current2}) it is inferred that the period of Aharonov-Bohm
oscillations in a ring of fractionally charged quasiparticles coincides
with the fundamental period ($\Phi_{0}$=hc/e) of oscillations in normal
eletron systems. This result \onlinecite{Kivelson,Bogachek.JETP} 
(see also Refs.~\onlinecite{Hatsugai} and \onlinecite{Gefen}) is a direct
consequence of  gauge invariance and follows in our calculations from
the fact that the wave function of fractional charge in a ring is multi-valued.
Below we show, however,  that the temperature dependence of the oscillation
amplitude is sensitive to the choice of statistics.
At first we examine Eqs.~(\ref{current2}) and (\ref{SC}) for $q>1$ in the low
temperature limit, $T \ll \Delta$. 
The low temperature behavior 
of the integrals (\ref{SC}) are
\begin{equation}
S_{j}^{(q)} \simeq \frac{1}{2 \pi^{2}} \frac{\beta \Delta}{j} \gg
C_{j}^{(q)} \simeq Const
\label{SC3}
\end{equation}
According to Eqs.~(\ref{current2}) and (\ref{SC3}) one then finds the persistent
current of fractionally charged exclusons in this limit to be
\begin{equation}
I_{q}(\Phi,T=0) = \frac{2}{\pi} I_{0} \sum_{j=1}^{\infty} \frac{
\sin \left(2 \pi j \frac{\Phi}{\Phi_{0}} \right)
\cos \left( 2 \pi j \frac{q \mu_{s}}{\Delta} \right) }{j},
\label{lowT}
\end{equation}
where $I_{0} = ev_{F}/L$. Notice that not only the period of oscillations,
but the amplitude , $I_{0}$, as well exactly coincides with the one for 
spinless fermions \onlinecite{Cheung}. A simple argument shows why the
persistent current  for exclusons does not depend on the statistical parameter
$\lambda$=1/q: The partial current carried by a quasiparticle 
$I_{s} = q_{s} v_{F} /L = I_{0}/q$ is q times smaller than that carried by an
electron, but since on average , $q$ exclusons occupy the state with a given
orbital  momentum, the total persistent current equals $I_{0}$.
Formally in Eq.~(\ref{lowT}), the statistical parameter $q$ enters the 
oscillating factor multiplied by the chemical potential $\mu_{s}$.
However, in the physical situation studied (a ring weakly connected
to an electron reservoir) only integer charges (electrons) can be
transported betwen ring and reservoir. Therefore
the controllable parameter in our case is not the chemical potential
of exclusons $\mu_{s}$, but the chemical potential of electrons in the
reservoir $\mu = q \mu_{s}$. In other words it is impossible to measure
fractional charge using Aharonov-Bohm oscillations at low temperatures.
Fractional charge is masked by fractional statistics.

Can we expand this claim to finite temperatures? Generally speaking --- no.
Anomalous (fractional) statistics in condensed matter can be conceived as
a result of an approximative (long wavelength) description of
specific interactions of particles with normal statistics.
At finite temperatures  thermally activated excitations
of strongly correlated systems contribute to the persistent current
and the characteristics of the current could depend on properties of the
interaction. The situation is similar in the presence of impurities violating
translational invariance of a system.
We show below that the temperature behavior of
Aharonov-Bohm oscillations due to quasiparticles
obeying generalized statistics is anomalous.

At finite temperatures it is impossible to obtain analytic expressions
for the integrals of Eq.~(\ref{SC}) at arbitrary statistical parameters.
As an example we calculate the persistent current for semions (q=2) in the
high temperature region, where it turns out to be possible. Other issues will be
addressed through numerical calculations.
For $q=2$ ($\lambda$=1/2) the algebraic equation (\ref{y}) is quadratic and
one gets a simple analytic expression for the distribution function.
In the case considered Eq.~(\ref{SC}) takes the form
\begin{eqnarray}
\left\{ \begin{array}{ll}
S_{j}^{(2)}  \\ 
C_{j}^{(2)} \end{array} 
\right\} &=&
-\frac{1}{\pi}
\left\{ \begin{array}{ll}
Im \\
Re \end{array} 
\right\}
\int_{-\infty}^{\infty} \frac{ dx e^{i \Omega_{j} x} } {
\sqrt{ \left( \frac{1}{2} \right)^{2} + e^{2 x}} } = \nonumber \\
&& - \frac{1}{\pi} 
\left\{ \begin{array}{ll}
Im \\
Re \end{array} 
\right\}
 \int_{C}\frac{ dz e^{i \Omega_{j} z} } {
\sqrt{ \left( \frac{1}{2} \right)^{2} + e^{2 z}} } ,
\end{eqnarray}
where $\Omega_{j} \equiv 4 \pi j/\beta \Delta$ and the integration contour
in the complex $z$-plane is shown in Fig.~1. The asymptotic behavior  
of the integrals at high temperatures is given by
\begin{eqnarray}
\left\{ \begin{array}{ll}
S_{j}^{(2)}  \\ 
C_{j}^{(2)} \end{array} 
\right\}
&\simeq&
\left\{ \begin{array}{ll} + \\ - \end{array} \right\}
\sqrt{\frac{2}{\pi^{2}} \frac{\beta \Delta}{j}}
\exp \left( -j \frac{2 \pi^{2}}{\beta \Delta} \right) \times \nonumber \\
&& \left\{ \begin{array}{ll}
\sin \\ \cos \end{array} \right\}
\left( \frac{4 \pi \ln 2}{\beta \Delta} + \frac{\pi}{4} \right)
\label{SChighT}
\end{eqnarray}


In contrast to the case of Fermi statistics, Eq.~(\ref{SC2}), both 
integrals in Eq. (\ref{SChighT}) are of the same order. Moreover,
they oscillate as a function of temperature with the period
$\Delta T_{2} = \Delta/2 \ln 2$. 
%

Fig. 2 demonstrates the variation of the oscillation amplitude
with temperature  for different values of the 
statistical parameter $\lambda = 1/q$ (at $\mu = k \Delta$, $k$ is an integer).
The crossover temperature 
$T^{*} = \Delta / 2 \pi^{2} = \hbar v_{F}/\pi L$ (see Eq.(\ref{SChighT}))
does not depend on the factor q.
It exactly coincides with the analogous temperature for a gas of noninteracting
electrons at fixed chemical potential \onlinecite{Kulik.BIL,Cheung}.
The period of temperature oscillations --- on the other hand --- does depend
(although only logarithmically) on the statistical factor
$\Delta T_{q} \sim \Delta / \ln q$. For large enough $q$ the oscillations
in question become appreciable already in the crossover region (see Fig.~2).


It was shown in Ref.~\onlinecite{WuYu} that at low temperatures an ideal
gas with the statistical parameter $\lambda$ can be described by the
Luttinger liquid model with the Haldane parameter equal to $\lambda$.
While doing this mapping we should take into account that the multivalued
nature of the wavefunction of fractionally charged particles mentioned above,
corresponds to introducing a number of vaccuum states and a corresponding
number of fields $\varphi_{k}$ describing excitations of the system on 
top of each vacuum state. The equivalence of both descriptions at zero
temperature mentioned in Ref.~\onlinecite{WuYu} can be proven in our
case if when calculating the persistent current one uses the following twisted 
boundary conditions for the fields $\varphi_{k}$ as a function of coordinate 
$x$ and imaginary time $\tau$:
\begin{equation}
\sum_{k=1}^{q} \left\{ \varphi_{k}(x,\tau+\beta) - \varphi_{k}(x,\tau) \right\}=
2 \pi n q .
\label{boundary}
\end{equation}
Here $n=0,\pm 1,\pm 2$,... is the winding number and $\beta=T^{-1}$.
By evaluating the partition function represented as a 
functional integral over the fields $\varphi_{k}$ described by a standard
quadratic Lagrangian in the long wavelength representation, 
one finds that the  persistent
current at zero temperature is identical to Eq.~(\ref{lowT}).
At the same time we can show that there are no oscillations of the persistent
current amplitude with temperature within the Luttinger liquid approach.
Therefore we conclude that the finite temperature behavior of the persistent
current calculated using the two approaches are different. This is not surprising
because the mapping of the excluson gas model to the Luttinger liquid model
\onlinecite{WuYu} is valid only at zero temperature.

In conclusion we have discussed the possibility to
detect effects of fractional charge (statistics) in low-dimensional
mesoscopic systems. 
The existence of fractional charge have been predicted by theory
for different systems in solid state physics
(see e.g. Ref.~\onlinecite{Krive.SovPhys}), but only two of them have been
studied systematically: quasi-1D metals with charge density waves
(see the review Ref.~\onlinecite{Gruner}) and the 2D electron gas in the
fractional quantum Hall effect (FQHE) regime \onlinecite{Prange}. It is known
that quasiparticle excitations (vortices) in the FQHE possess fractional charge
and statistics \onlinecite{Laughlin,Halperin}. Although this prediction has still
not been verified directly in an experiment, it is based on the well developed
and experimentally confirmed theory of the FQHE.

The situation in quasi-1D Pierles-Fr\"ohlich systems is more intricate.
At present there is no unambiguous experimental
evidence for the existence of fractional charged solitons in 1D channels. 
We would like to suggest that it may well happen that experiments in mesoscopic
systems (for example in quantum wires) could be more suitable for
measuring fractional charge and -statistics.
Under usual conditions, current carrying excitations in 1D quantum wires
are quasiparticles with integer charge. However, it is possible to
conceive an experimental situation when the charge motion along the channel
occurs in the presence of a periodic external potential.
If the average density $n_{e} = 1/a$ of electrons in the channel is close
to commensurate fillings $n_{e}a_{p} = q/p$ ($a_{p}$ is the period of
the channel modulation, $q>p$ are integers) the electron system
will be near the commensurate-incommensurate phase
transition. In this case the free charge carriers are solitons with fractional
charge $q_{s} = e/q$ \onlinecite{Kolomeisky}.
The persistent current of a soliton ring in the metallic
(incommensurate) phase then should exhibit the anomalous features studied 
in the present paper.

We are grateful to S. M. Girvin for many helpful discussions.
I.K. acknowledges the hospitality of the Department of 
Applied Physics, CTH/GU. This work was supported by the Swedish Royal Academy
of Science, the Swedish Natural Science Research Council, INTAS grant 94-3962,
and by grant U2K200 from the Joint Fund of the Gov. of Ukraine and
the Intern. Science Foundation.

\begin{figure}
\caption{
The  integration contour in the complex plane for the analytically solvable
case $\lambda = 1/2$. 
}
\end{figure}

\begin{figure}
\caption{
Temperature dependence of the persistent current calculated 
for two different statistical parameters 
$\lambda = 1/q$.
Note that the current for $q=3$ has been magnified by a factor of 100.
While the crossover temperature does not depend on $\lambda$, the
``period'' of nonmonotonic temperature behavior do so as can be seen
from the figure.
}
\end{figure}

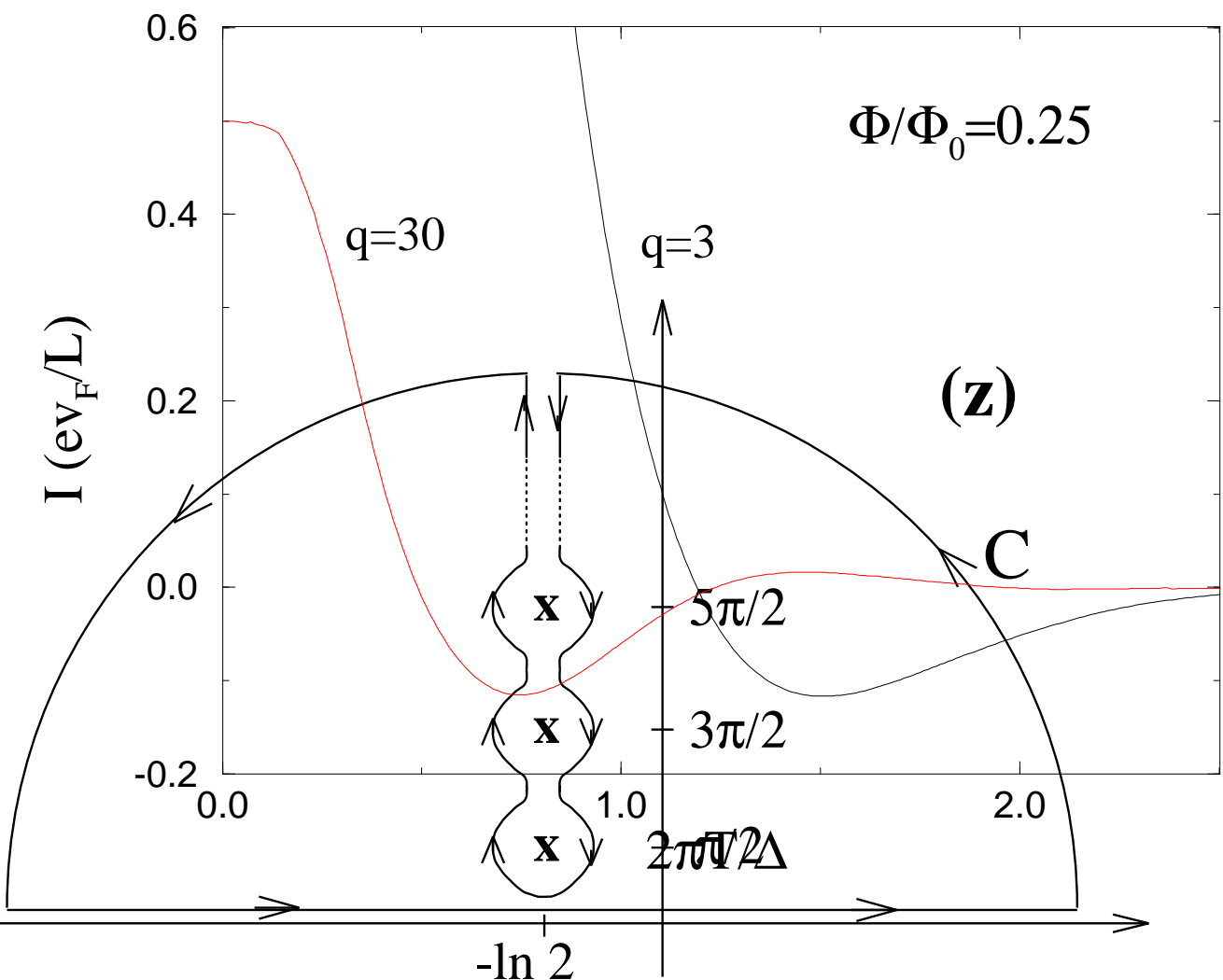
\end{document}